# Search for a new charged particle in the mass range of 2-100 *MeV*.

M.H. Anikina [1], V. A. Nikitin [1], V.C. Rikhvitsky[1]
[1] *Joint Institute for Nuclear Research. Dubna.*

**Поиск новой заряженной частицы в интервале массы 2 – 100 *МэВ*.**

*М.Х. Аникина [1], В.А. Никитин [1], В.С. Рихвицкий[1].*
[1]*Объединённый институт ядерных исследований. Дубна.*

## Summary.

В теории электрослабых взаимодействий отсутствует запрет на существование частиц с массой отличной от массы электрона, мюона и тауона. Ставится задача выполнить поиск новой частицы в диапазоне массы 2 – 100 *МэВ*. Поиск выполнен на фото материале 2-м пропановой пузырьковой камеры. Камера экспонирована в пучке протонов 10 *ГэВ* на Синхрофазотроне ОИЯИ. Просмотрено ~55 тыс. стереофотографий. Анализируются события конверсии γ кванта в пару заряженных частиц. Найдены 50 аномальных событий, в которых одна частица пары останавливается в объёме камеры, имеет в конце пробега повышенную плотность трека и при идентификации обнаруживает массу ~*8 МэВ*. При этом среднее значение массы новой частицы составляет (8,5 ± 2,5) *МэВ*.

In the theory of electroweak interactions, there is no prohibition on the existence of particles with a mass other than that of an electron, muon and tauon. The task is to search for a new particle in the mass range of 2-100 *MeV*. The search was performed using the photo material of the 2-m propane bubble chamber. The chamber has been exposed in a 10 *GeV* proton beam at the JINR Synchrophasotron. For its search, ~ 55 thousand stereo photos have been scanned. The events of the *γ* quantum conversion into a pair of charged particles have been analyzed. 50 anomalous events have been found in which one particle of the pair stops in the volume of the chamber, has an increased track density at the end of the range and, upon identification, detects a mass of ~ 8 *MeV*. The average value of the mass of the new particle is (8.5 ± 2.5) *MeV*.

## Introduction.

In the theory of electroweak interactions, there is no prohibition on the existence of particles with a mass other than that of an electron, muon and tauon. The motivation for the search for new particles is some anomalies in the scattering of electrons near radioactive sources. There is also a deviation in the shape of atmospheric cosmic showers from the calculation result. However, these indications are indirect. They cannot be considered reliable evidence of the existence of new particles.

In the period of 1960 - 1972, a search for unknown charged particles [1 - 6] was performed at various laboratories.

Paper[1] describes an experiment to find out whether gamma quanta generate particles with a mass from 6 to 25 electron masses. The study was performed at the FIAN synchrotron (Moscow) on a gamma ray beam with the maximum energy of 265 *MeV*. Charged particles were formed in the lead target in the $\gamma \to l + l$ conversion reaction. The momentum of the particles was determined in the magnetic spectrometer, where their velocity was also measured by the time of flight between scintillation counters. The mass of the particle was calculated from the measured momentum and velocity. The desired particles were not registered. For example, for a mass $m = 10\ m_e$, the result is given: the calculated theoretical value of the cross section is $\sigma_{theor} = 110$ *mb*, the measured upper limit of the cross section is 4 *mb*. A similar result, but obtained with a different technique, is reported in paper[2]. The measurements were performed at the charged particle energy of 93 *MeV*. The search result is reported: the required mass is 10 $m_e$, the calculated cross section is $\sigma_{theor}=40$ *mb*, the measured upper bound of the cross section is $\sigma_{exp}=0.7$ *mb*. Paper[3] was performed at Stanford on a linear electron accelerator also on a beam of bremsstrahlung photons. The magnetic spectrometer generates a beam of conversion charged particles with a momentum $p = 300$ *MeV/c*. Particles from the spectrometer are sent to a scintillation telescope of 6 counters layered with lead plates. Muons have a range of 154 $g/cm^2$. In the range of $L>154$ $g/cm^2$, only particles with a mass less than the mass of the muon can appear. The measurement has shown that the countig rate in the area of $L>154$ $g/cm^2$ is 2 - 3 orders of magnitude lower than in the area of $L<154$ $g/cm^2$.

The authors conclude that the experiment excludes the existence of particles with masses in the range of 5-175 $m_e$ and with a production cross section following from the electro–magnetic theory. The permissible upper limit of the production cross section of an unknown lepton is not specified. In papers [4, 5, 6], the search is carried out in the region above the mass of the muon. The result is also negative.

## 1. Search for a new particle.

*a) Film scanning.*

The search for a new particle was performed on the photographic material of the 2 - m propane bubble chamber. In the 60s of the last century, the chamber was irradiated with protons having the energy of 10 $GeV$ at the Synchrophasotron of the JINR LHE. The size of the bubble chamber area, which can be observed in a stereo lens and on a viewing table, is 105 x 60 x 40 $cm^3$.

The chamber is located in the magnetic field with the strength of 1.5 $Tl$. The density of liquid propane is 0.43 $g/cm^3$. The radiation length of propane is $\lambda_{rad} = 104$ $cm$. The proton beam interacts with the nucleons of the propane molecule. As a result, multiple particle production takes place. These particles are mainly nucleons and pions. Neutral pions decay into gamma quanta, which convert a pair of charged particles into propane. They are mainly pairs of $e^+$, $e^-$.

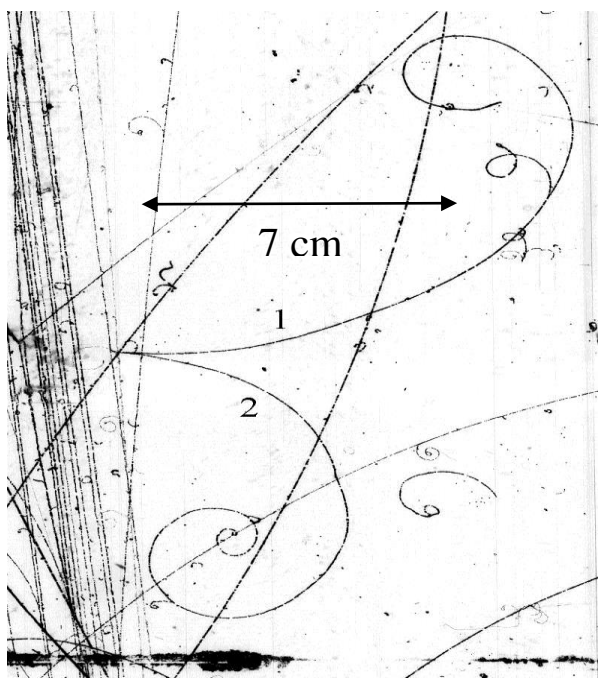 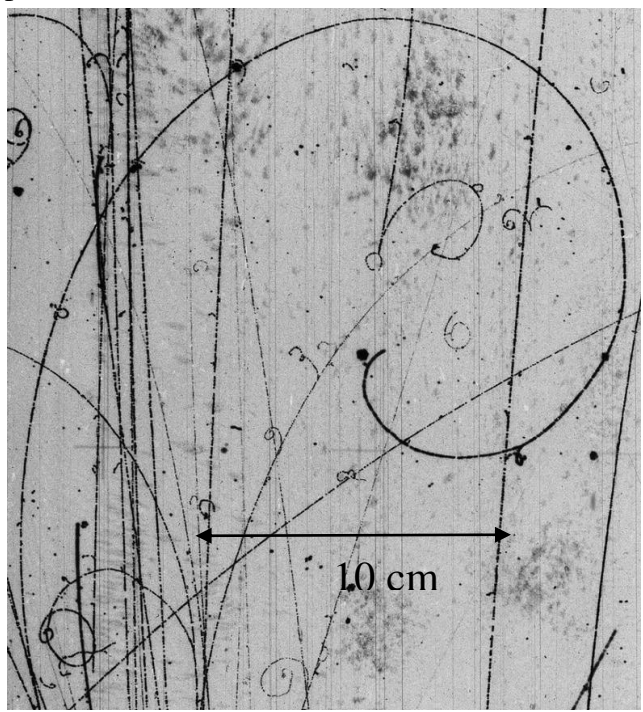

*Fig. 1.   1. The track of the negative particle with a stop in the chamber and an increased density at the end.   2. The track of a positive particle (positron).*

*Fig. 2.   The track of a positive particle with a stop in the chamber and an increased density at the end.*

55 thousand stereo photos have been viewed. When viewing the photographic material, the events of particle production pairs by gamma quanta $\gamma \rightarrow l + l$ have been selected, in which at least one particle remains in the viewed volume and has the increased optical density (fusion of bubbles on the track) near the last visible point. Attention is drawn to the curvature of the particle trajectory. It should be more than that of a muon and less than that of an electron. This criterion gives preference to particles in the mass range of 2-100 MeV. The particles of interest are with a mass greater than the mass of an electron. For the desired particle ionization increases by ~ 1.5 – 2.0 times at the end of the range at a distance of 2 - 4 *cm* or less from the end point. For comparison, in the electron, this increase of ionization has been observed at a distance of 2 *mm* or less from the end point. That is, the electron remains a particle with minimal ionization along almost the entire visible trajectory. With the increase of ionization losses, the number of bubbles formed on the trajectory increases. When viewing the tracks, the gaps disappear at the end and the bubbles merge together into a solid line.

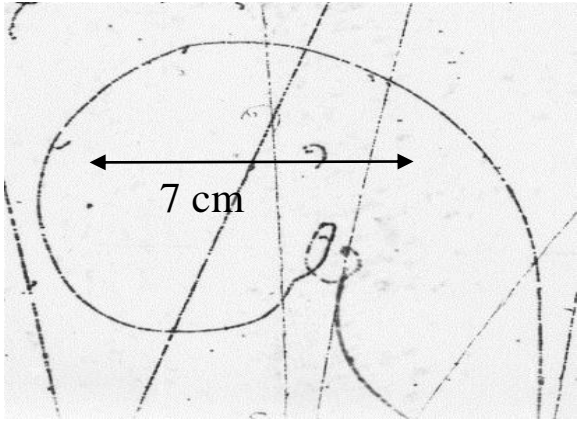
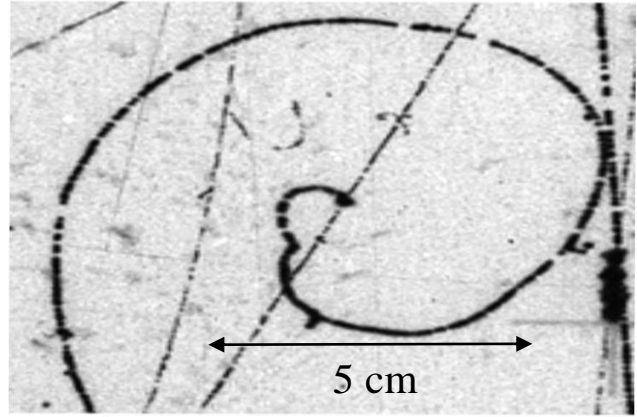

*Fig. 3. The track of a negative particle with an increased density at the end. A negative particle with minimal ionization (electron) is emitted.*

*Fig. 4. The track of a positive particle with an increased density at the end. A positive particle with minimal ionization (positron) is emitted.*

The characteristic tracks selected to be measured while viewing are shown in Figs. 1 - 4. We are not interested in the second track of the fork if it does not meet the selection criteria. The analysis procedure consists in measuring the spatial coordinates of the track points and obtaining the main characteristics of the particle: mass, depth angle and ionization losses.

*b) Restoration of the trajectory and calculation of the particle mass.*

The track of the slow particle is digitized in three photo images. Coordinates of ~ 10 - 40 points are measured on it. The trajectory is restored in three-dimensional space [8]. It is divided into several intervals. The length of the interval is set by the number of points measured on it: 5, 7 or 9 points. At each interval, the curvature of the trajectory is calculated from three points repeatedly with every possible combination of three points. The resulting set of radii of the curvature is approximated by a polynomial. As a result, the radius of the curvature $R_{exp}(l)$ is calculated at each point in the plane normal to the vector of the magnetic field (camera plane). Here *l* is the residual range – the distance from the end of the track to a given point in three-dimensional space. Fig. 5 shows the data for a typical track. Panel 1 shows particle trajectory in chamber plane perpendicular to magnetic field. Panel 2 presents particle deep angle. As one can see, the measured trajectory curvature radius (panel 3) experiences a fluctuation due to the scattering of a particle on propane. The increased errors are attributed to this group and the analysis is repeated. Therefore, a semi-empirical function $R(l)$ approximate the experimental data - a solid red line on panel 3. Function $R(l)$ is used to calculate the particle momentum modulus: $p=k \cdot B \cdot R(l)/Cos(\alpha)$, where $\alpha$ is the angle between the vector tangent to the track and the camera plane - the depth angle, panel 2. *B* is the vertical projection of the magnetic field within the measured section of the track (*B* - *kGaus*, *R* – *cm*, *k* =0.3). The mass of the particle *m* is calculated at each point of the trajectory by solving the following equation:

$$T = \sqrt{p^2 + m^2} - m. \qquad (1)$$

Here *T* is the kinetic energy determined by the range *l*:

$$l(T, m) = c1 \frac{T^2}{T^{c2} + m} \qquad (2)$$

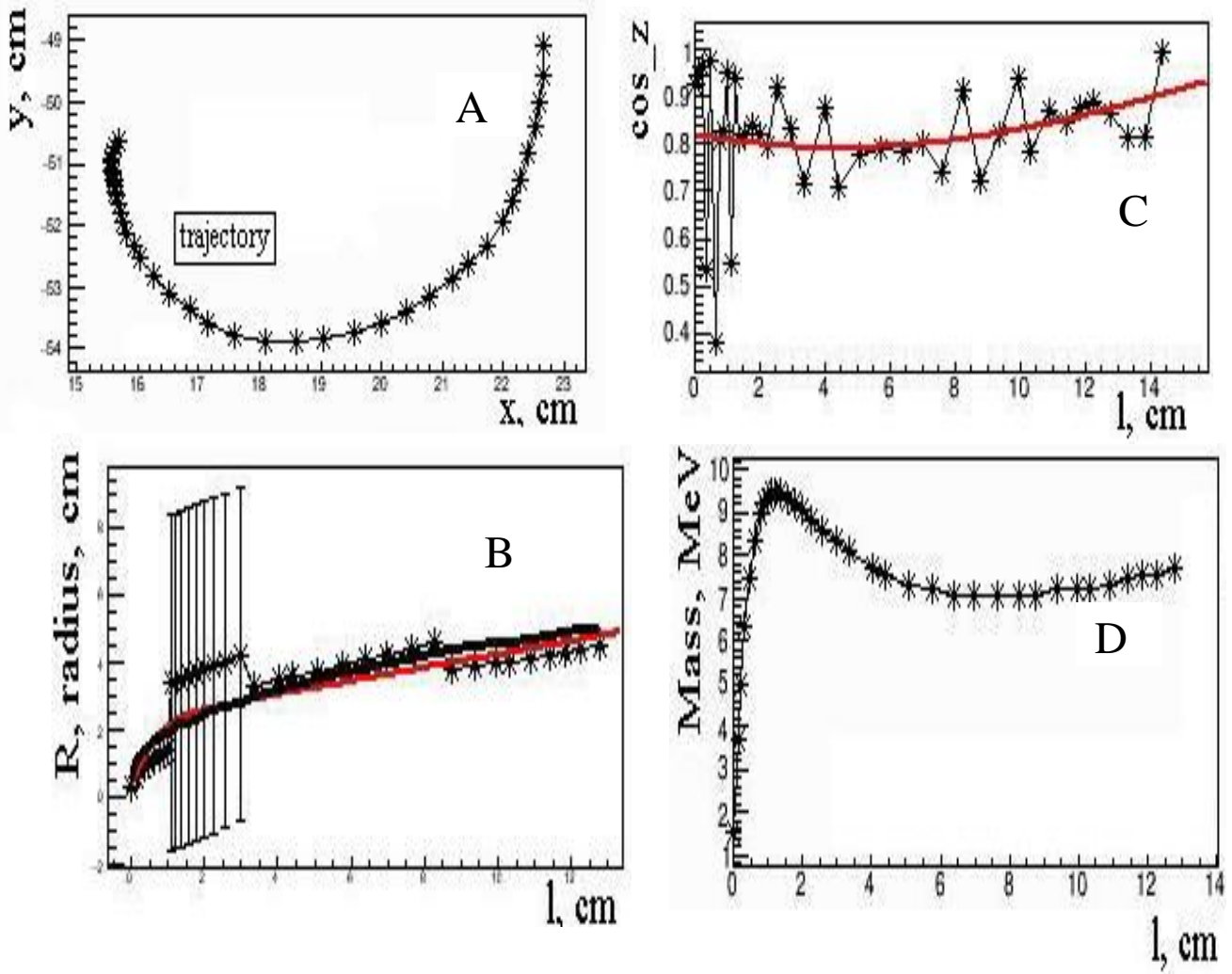

*Fig. 5. Features of the unknown negative particle with a mass of (7.6±1.1) MeV.
A) The trajectory of the particle. ). B) The radius of the curvature of the trajectory.
C) The cosine of the depth angle.    D) The mass of the particle calculated at each point
of the trajectory.*

      The semi empirical function (2) is derived by integrating the ionization loss function $dT/dl(l)$. The parameter $c1$ is determined by the known value of the minimum ionization value: $c1 \approx$ $(1.15+0.006*m) = 1/(dT/dl)_{min}$. In this form, formula (2) is applicable to analyze the particles with the masses from an electron to a pion. The parameter $c2=1.08$ approximately takes into account the relativistic increase in ionization with increase in the energy (range) of the particle. This mode becomes noticeable in the region of the particle's energy above its mass, i.e. $T>m$. The mass of a particle is defined as the weighted average value along the trajectory on the panel D. The systematic error follows from the variation of the value $m(l)$ along the track of the particle. Finally, the mass is calculated as the arithmetic mean of all estimates. The algorithm to determine the mass of particles is tested on groups of particles with a known mass (control tracks). These are the tracks of electrons and positrons modeled by the Geant-4 program - Fig.6.  These are the measured delta electrons, muons, and pions –   Figs. 7, 8, 9. Electrons and muons are visually identified: electron trajectories along the entire length have shown minimal ionization, and at the end, as a rule, have loops. The muon is visibly located by the decay to an electron. The pion is located by the decay to a muon and decay of a muon to an electron. All possible processes of distortion of the particle trajectory (multiple scattering, bremsstrahlung radiation, etc.) also take place for the trajectories of control tracks. The histograms obtained for the control tracks have shown the reliability of the applied algorithm to calculate the mass.

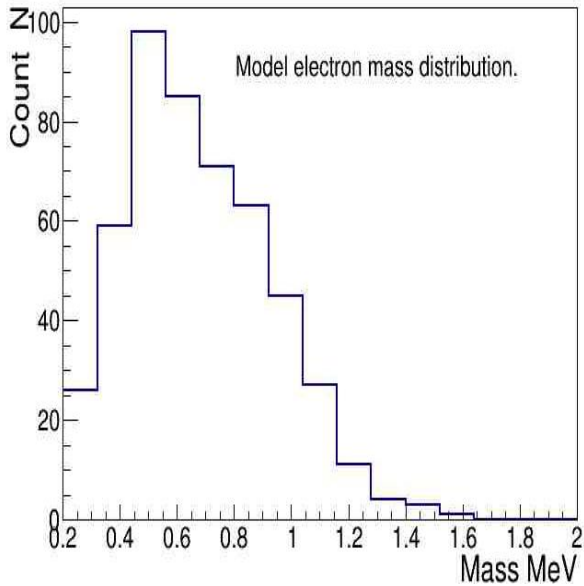

*Fig. 6. Electron mass spectrum modeled by the Geant-4 program*

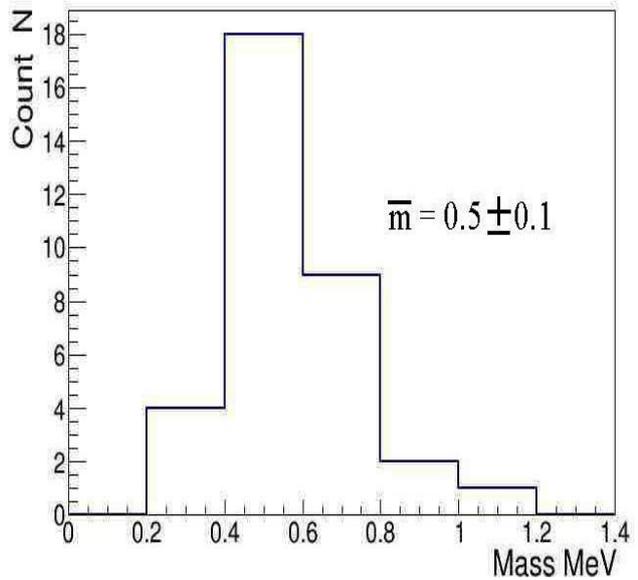

*Fig. 7. Mass spectrum of the measured delta electrons*

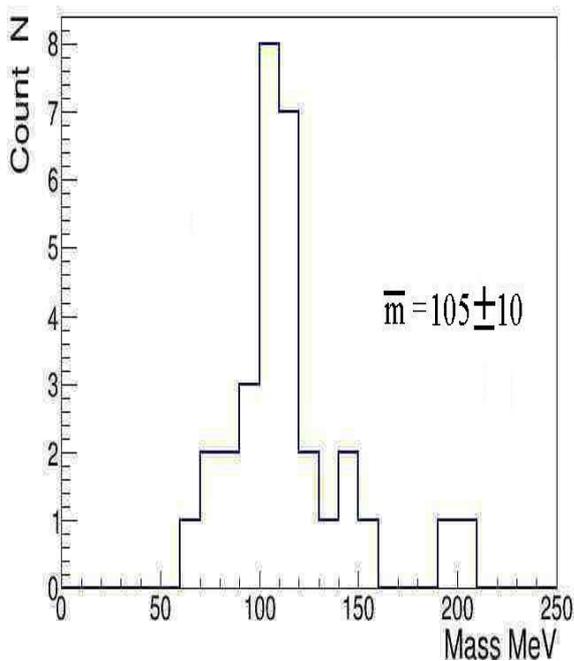

*Fig. 8. Mass spectrum of the measured muons*

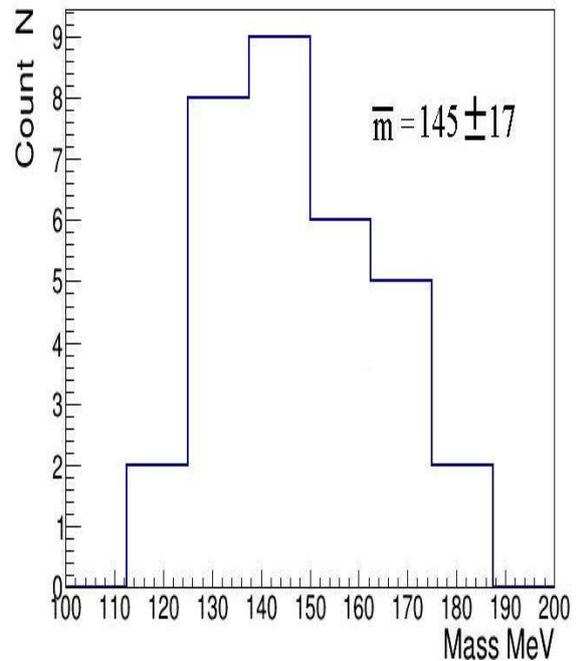

*Fig. 9. The mass spectrum of the measured pions*

*c) Determination of ionization losses.*

The method of searching for a new particle in this experiment consists in selecting tracks with increased ionization at the final site. When solving this problem, greatest concern is caused by possible instability of the chamer mode, which can distort the character of the track around the stop point. It is possible to have the variation of the lighting within the volume of the camera. The scanning has shown that in any illumination mode, the lumen between the bubbles or their fusion is visible reliably. This is visible on all projections, although the tracks have different density values on three different frames. It is due to the fact that the camera sees the track from different angles. Perhaps, there is a dependence of the degree of blackening of the track on the depth of its position in the chamber. To put it simply, the increased blackening near the glass is a consequence of the illumination mode. This dependence has been studied on dozens of tracks with minimal ionization. It has been found that there is dependence, but it appears only at the depth of 20-22 cm or more. Our measurements lay higher. We see the compaction of tracks before stopping with all changes in chamber modes and throughout its entire volume. This can be seen on the muon tracks having a compaction caused by ionization losses before decaying into an electron (Fig. 10). This indicates that the chamber reliably records the change in ionization along the particle track.

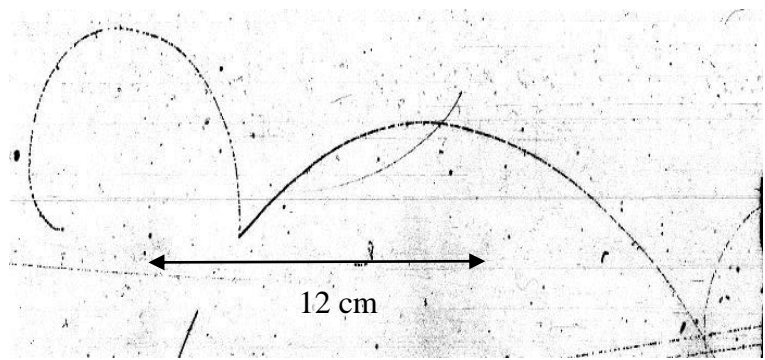

*Fig.10. The track of a muon emitting a decay electron. The increased density of track bubbles in front of the stop and decay point is clearly visible.*

*d) Measurement of blackening the particle track.*

To test the visual method of isolating particles with an increased degree of blackening, a program to quantify the amount of track blackening has been developed [8]. It is concluded in the following. The digitized stereo projections give access to individual pixels of the image. Therefore, it is possible to operate with the transparency of a single pixel. Transparency is characterized by a number in the range of $0 - 255$. It becomes available to determine the blackening of the track in each small interval. The dimension of pixel is 0.2 mm. Blackening is calculated in a photo where the track trajectory is free from the background. Sections of the track with a length of 30 pixels are measured sequentially. The blackening of each interval is normalized by the amount of blackening of the trajectory section with the gaps between the bubbles.

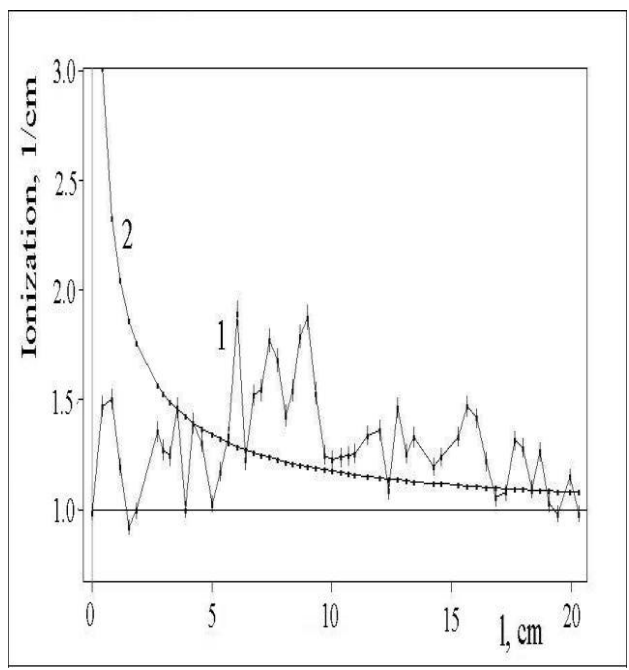
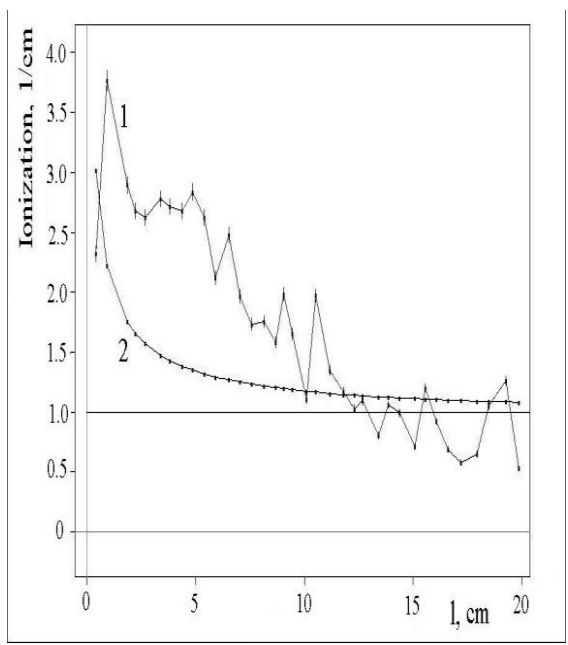

*Fig. 11. 1. Dependence of the relative blackening of the track on the range for a particle with minimal ionization. The particle enters the glass. 2. Calculated value of relative ionization losses for a particle with a mass of 8 MeV.*

*Fig. 12. 1. The dependence of the relative blackening of the track on the range for µ meson. 2. The calculated value of relative ionization losses for a particle with a mass of 8 MeV.*

Graphs of the degree of relative blackening for three different particles are shown in Fig. 11 - a track with minimal ionization, Fig. 12 - a muon track with fused bubbles at a length of about 10 *cm* and Fig. 13 - a track with fused bubbles at a length of $2 - 3$ *cm*. The spread-out of blackening values of different points is associated with fluctuations in the density of bubbles and gaps in a circle of 30 pixels designed to measure blackening for one selected point. The graphs illustrate the calculated curve of relative ionization losses for the particle with a mass of 8 *MeV*. It can be seen that only for the track in in Fig. 13, the calculated and experimental curves coincide. Thus, this is the track of a particle with a mass of ~ 8 *MeV*.

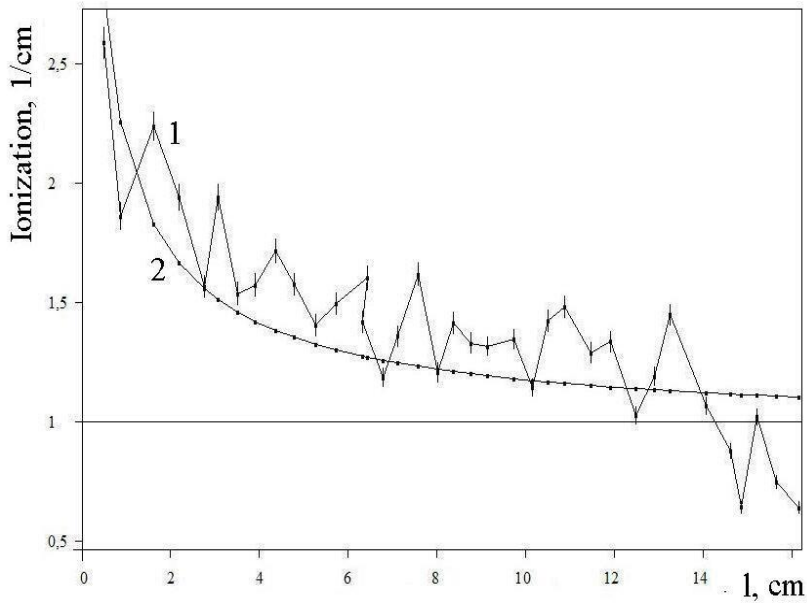

*Fig. 13. 1. Dependence of the relative blackening of the track on the range unknown particle. 2. The calculated value of relative ionization losses for a particle with mass 8 MeV.*

This program has been successfully used to verify the existence of the blackness dependence on the depth of the track in the camera (see above). Measurements on the tracks with a solid end have confirmed the reliability of the visual determination of increased ionization criterion.

*e) Tracks going into the camera glass.*

We have included the tracks whose stopping point lies in the camera glass. This increases the error in determining the mass, since the measured stop point is offset from the original one. A study of possible errors was made when including these tracks into the number of candidates of a new particle. Fig. 11 shows that the particle with minimal ionization going into the upper glass of the chamber does not detect blackening growth on its trajectory. I.e., the selection of tracks with blackening near the glass cannot be explained by the peculiarities of lighting in the chamber. The increased blackening of the track indicates the proximity of the end point. Then the range can be estimated with negligible definiteness. To verify this assumption, several dozen tracks stopped in the camera have been examined.

The masses of the corresponding particles have been measured in two versions: on the full visible trajectory and on the trajectory with the cut end. The length of the cut end is 1, 2, 3, 4 *cm*. It is shown that the mass in case of the cutting end exceeds the correct mass by 0.5 – 2.0 *MeV*, which falls within the limit of statistical and systematic errors. This estimate justifies the inclusion of the tracks with a stop in the glass in the statistics of the desired events, although it may lead to a shift in the mass estimate by an average of 1 *MeV* to the right.

We also note that the number of negative particles in about 200 selected tracks is three times less than the number of the positive ones. This is explained by the asymmetric position of the beam of protons in the chamber. It is shifted to one edge, and the negative particles more often go into the side wall.

*f) Background particles.*

There is a large number of particles on the viewed films, which are called background, because they can simulate the desired particles. They are electrons and positrons. A positron, propagating in propane, can annihilate on an electron. In this case, its apparent trajectory is shortened as a result of annihilation. The calculated mass of the positron in such an event turns out to be overestimated. And this leads to a dangerous systematic error in determining the mass of the desired unknown particle. An electron can also shorten the trajectory length by emitting photons bremsstrahlung. The mass values occupy the entire range of interest.

Fig. 14 shows a histogram of the mass of these particles. On a randomly selected hundred frames, all the particles with a visible end in the camera and with minimal ionization, i. e. with gaps between the bubbles along the entire trajectory, were measured. We have got rid of the main contribution of these background particles even at the scanning stage. The histogram in Fig. 15 includes all the tracks found on the array of 55,000 frames with the given feature (150 particles in total). It can be seen that in the histogram of the mass of the tracks selected by the solid end of the track, there is a peak at a mass of 8 MeV which appears immediately. But this spectrum (Fig. 15) still contains some background particles. These particles experience scattering close to the end of the track and receive a large depth angle, which causes a visible increased compaction of the track. But after measuring the depth angle of the trajectory, the above tracks are removed according to the criterion of limiting the depth angle.

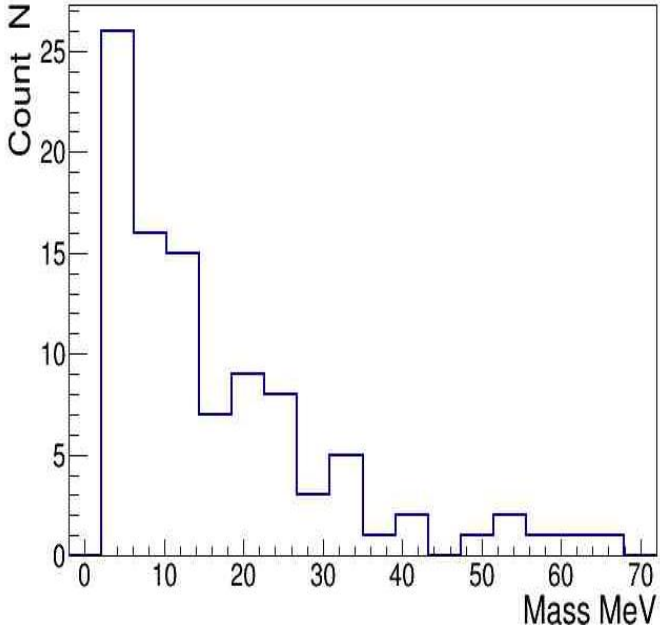
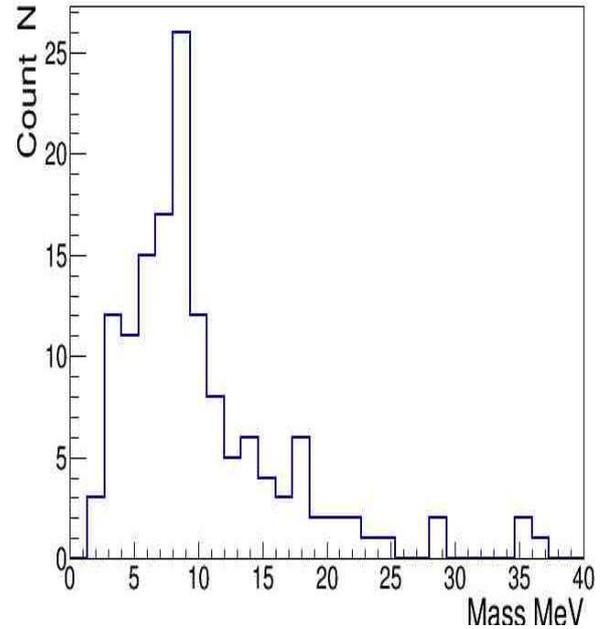

*Fig. 14. The mass spectrum of particles with minimal ionization having the last visible track point in the camera (background particles).*

*Fig. 15. The mass spectrum of particles with the increased density of bubbles at the end of the track and with any deep angle (a mixture of background and unknown particles).*

The study of the dependence of the process of visible bubble fusion on the slope has shown that at the angles less than 40 degrees the gaps between the bubbles are always visible for particles with minimal ionization, and at the inclination of more than 50 degrees the bubbles merge into a solid line. Thus, it is certain that for the tracks with an inclination angle of less than 40 degrees, the union of bubbles at the end indicates the increased ionization of the particle. Then the last visible point of the track trajectory is the true stop point and the calculated track mass is correct. Some of the unusual tracks stopping in the chamber emit a low energy particle with the identical charge sign and minimal ionization at the end (Figs. 3, 4). Electrons and positrons are emitted. There are no radiated particles in the tracks going into the glass. They get lost in the glass. The emission of positrons by positively charged unusual particles rejects the assumption that radiation is simulated by scattering a particle with electrons.

The work [7] was performed by analyzing the particles detected in the same propane bubble chamber as presented in this publicanion. The chamber has been irradiated at the JINR Synchrophasotron with 10 GeV/$c^2$ protons. The author has developed a new particle identification algorithm. An analytical formula (hadroid) describing the trajectory of particle propagation in a substance has obtained. The formula includes the mass of the particle as an unknown desired parameter. It is found using the least squares method when describing the trajectory. It is assumed that the last point of the trajectory is the stopping point of the particle. 1,474 particles were taken for identification. These are mainly positrons, muons, pions, protons. Some of the positrons are annihilated by electrons. It is known that tracks with a lost end give an overestimated (false) mass value when analyzed. The author observes an almost uniform mass spectrum in the interval between an electron and a muon. In the events of the $\gamma \rightarrow l + l$ conversion, one of the particles may have a mass higher than the mass of the electron (the desired new particle). However, such events are rare and for their selection, it is necessary to apply the criterion of increased ionization at the end of the track (see below). In [7], tracks taken without pre-selection are analyzed. Therefore, the task of searching for a new particle has not been achieved.

### 3. The mass spectrum of the unknown particle.

50 events were registered in which the stopped particles satisfy the following selection rules (signs of the unknown particle).

a) The particle stops in the upper half of the chamber.

b) At the end of the track (3-4 *cm* from the stop point), the particle exhibits increased ionization, i.e. the bubbles merge into a solid line.

c) The depth angle does not exceed 40 degrees over the entire part of the trajectory with increased density.

d) Trajectory curvature is more than that of a muon and less than that of an electron.

e) Tracks with gaps between bubbles on the entire trajectory are excluded as background particles. Therefore, it is assumed that the background in our final result (Figs. 16, 18) is negligible. The average value of the particle mass represented by the distribution Fig. 16 is $(8.5 \pm 2.5)$ *MeV*.

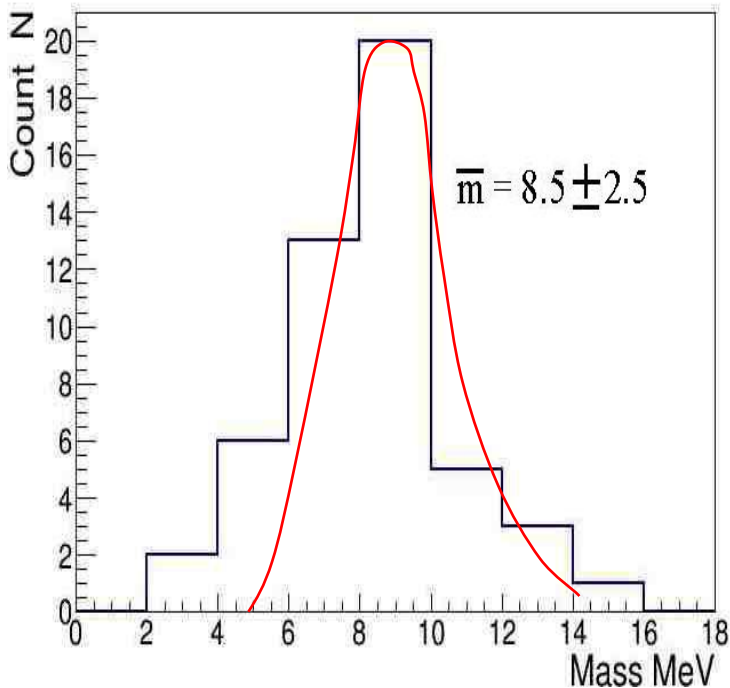 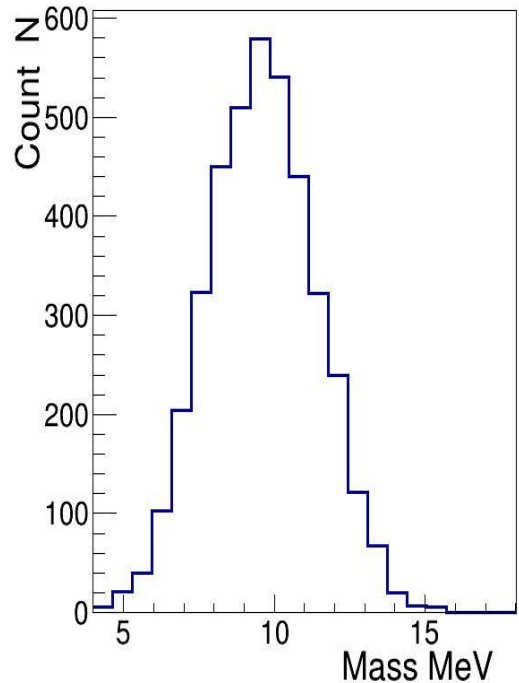

*Fig. 16. 1. Mass spectrum of particles satisfy the selection rules (signs of the unknown particle). 2. A copy of the simulated mass spectrum, Fig. 17.*

*Fig. 17. 1. Mass spectrum of simulated particles with a mass of 8 MeV.*

Fig. 18 shows the spectrum of the unknown particle in the full studied range.

The tracks of unknown particles having a large inclination angle, thus, remain beyond the opportunity of reliable identification (Fig. 19). They are not included in the final result.

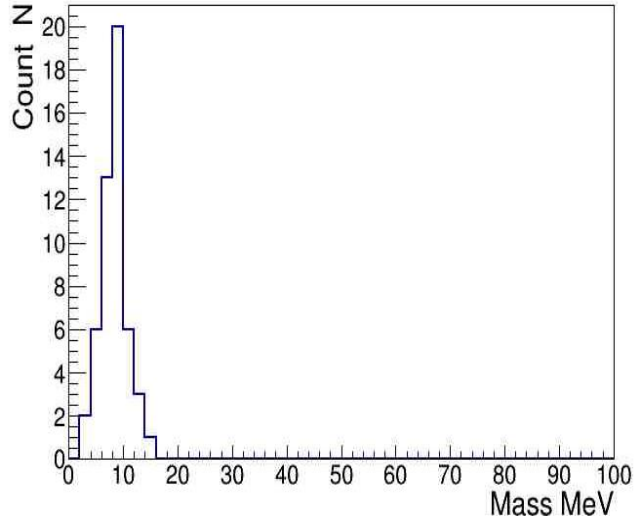 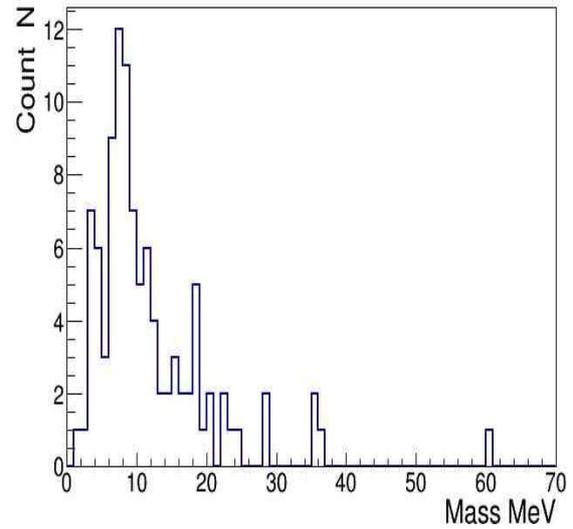

*Fig. 18. The same as in Fig. 16, but the total observed mass range is presented.*

*Fig. 19. The mass spectrum of particles with the increased density at the end, having the inclination angle of more than 40 degrees.*

1. **4. Estimation of the lifetime of the unknown particle.**

In some cases, an electron or positron is observed at the end of the track (Figs. 4, 5), which indicates the decay of the unknown particle. The electron energy in these events is ~ 0.5-1.5 MeV. The sensitivity time of the camera is ~ 10 *ms*. Some of the unknown particles manage to disintegrate during this time. This allows us to estimate the lifetime of the unknown particle: ~ 10 – 30 *ms*. Considering that the particle mass is ~ 8 MeV, the observed decay channel indicates the presence of one or more neutral particles among the decay products.

**5. Estimation of the production cross section of the unknown particle.**

Let's estimate an important value – the production cross section of the unknown particle in the reaction $\gamma \rightarrow l + l$ on the nucleus of the propane molecule ($C_3H_8$). The number of nucleus on the length of the chamber is $N_N = 2.6 \cdot 10^{24}$ $cm^{-2}$. The number of primary protons trapped in the camera on the entire viewed photographic material $I = 5.5 \times 10^5$. The number of the found unknown particles is $N_{alpart}=50$ in the momentum range of 20-150 *MeV/c*. The efficiency of registering the unknown particle is estimated by taking three parameters: the depth angle of the particle trajectory $\alpha \leq 40$ *deg.*, the average length of the photon trajectory in the chamber $L_\gamma = 60$ *cm*, the radiation length of propane $L_{rad} = 104$ *cm*. We have obtained the estimation of the efficiency of this unknown particle (we call it anomalon) registration:

$$f = (2 \cdot \alpha/\pi)(1-\exp(-L_\gamma/L_{rad})) = 0{,}4*0{,}42 = 0{,}17.$$

The desired lower bound of the cross section $\sigma_{alpart}$ is calculated using the following formula:

$$N_{alpart} = I \cdot N_N \cdot f \cdot \sigma_{alpart}.$$

Result: $\sigma_{alpart} = (0.19 \pm 0.03)$ *mb*. This value lies below the previously published values of the upper bound of the new particle production cross section.

## 6. The nature of unknown particle.

The assumption that a pair of particles made in the conversion of a gamma quantum, one of which is the new one, was verified in [9]. The kinematics of the reaction $\gamma+A \to l + l + A$ has been also verified. Two reaction versions are considered: $A$ is the carbon nucleus, or it is the proton. The momenta of the particles $l, l$ have been measured. The masses for the two leptons are set to 0.5 and 8 $MeV$. The momentum of the recoil nucleus $A$ varies in magnitude and angle. A solution has been searched for where the recoil nucleus range does not exceed the radius of the bubble. We state that in no case the track of A was registered at the conversion point. Satisfactory solutions were found for all forks with new particle (abbreviated as *a)*. The pair (*l, a*) is converted by a photon (Fig. 1), whose aromatic number is 0, therefore, the particle, *a,* (new one) has the aromatic number of an electron with the opposite sign. I.e., it does not have its own new aromatic number in analogy with leptons μ, τ. Therefore, it is not related to the new (fourth) type of neutrino. Otherwise, the nature of particle *a* is unknown.

### Conclusion

In the standard particle model, there is no prohibition on the existence of leptons with a mass other than that of an electron, muon, and tauon. The particle with a mass in the range of 2-100 *MeV* was searched for in the photo material of the 2-m propane bubble chamber. In the 60s of the last century, the chamber was irradiated with protons having the energy of 10 *GeV* at the Synchrophasotron of the JINR. To solve the formulated task, 55 thousand stereo photographs were viewed. The selected photos were digitized by scanning. The program to obtain spatial coordinates of points on the trajectory of a new particle has been improved. Programs have been written to measure the coordinates of points on a computer and blacken the trajectory of particles. The program has been written to calculate the mass of stopped particles using the curvature and range at each point of the trajectory. The simulation of the propagation of electrons, positrons, muons and unknown particles in propane using the GEANT-4 program has been accomplished.

The events of the conversion of a gamma quantum into a pair of charged particles have been analyzed. 50 anomalous events were found in which one component of the pair stops in the chamber volume and, upon identification, registers a mass of ~ 8 *MeV*. The average mass of the new particle is (8.5 ± 2.5) *MeV*. Several events of the new particle production paired with an electron or positron have been observed. Consequently, the lepton number of the new particle coincides with the lepton number of an electron or positron. The chamber allows one to observe the new particles in the momentum interval of 20-150 *MeV/c*. In the photo some of the new particles stopped in propane have shown the image of decay into an electron or positron. The lifetime of the new particle should be comparable to the sensitivity time of the bubble chamber, i.e. about 10-30 *ms*. The lower bound of the cross section production of the new particle in proton-nuclei interactions has been determined. It is (0.19±0.03) *mb*. This value lies below the previously published values of the upper boundary of the new particle production cross section.

Thus, this study indicates the existence of a previously unknown particle.

The authors are grateful to A.A. Baldin and A.Yu. Troyan for the opportunity of working with the photo material of the propane chamber, to A.V. Beloborodov - for organizing and maintaining the database of digitized films. We are grateful to A.V. Belyaev for his assistance and help in the work: for analyzing digitized images of the desired events and restoring the spatial coordinates of points on the particle track. He has also developed a track blackening measurement program. We are grateful to Alexander Malakhov for his advices and valuable recommendations. We thank P. V. Nomokonov and Yu.P. Petukhov for modeling the processes of particle propagation in propane by means of the GEANT-4 program. Laboratory assistants Asmik Grigoryan, Elena Dmitrieva and Tatyana Borisova viewed a large volume of photographic material, selecting the desired events. They also performed a scan of the selected photos, for which we are very grateful to them.